\documentstyle[12pt]{article}

\renewcommand{\baselinestretch}{1.2}

\def \Ref#1{[\,#1\,]}
\def \sect#1{\bigskip\vbox{\bigskip\leftline{\bf #1}}\vskip1ex\nobreak}

\def \R#1{\hbox{\bf #1}}
\def \degree{\hbox{$^{\circ}$}}
\def \imply{\hbox{$\Rightarrow$}}
\def \Kbar{\hbox{$\bar{K}$}}

\begin{document}

\pagestyle{empty} 

\rightline{UTAPHY-HEP-3}
\rightline{\today}

\begin{center}

NONET SYMMETRY AND TWO-BODY DECAYS OF CHARMED MESONS
\vskip 8ex

Waikwok Kwong and S. P. Rosen

{\it Department of Physics, The University of Texas at Arlington, \\
     Arlington, Texas 76019}

\end{center}

\vskip 8ex

\centerline{\it ABSTRACT}
\begin{quote}
The decay of charmed mesons into pseudoscalar (P) and vector (V) mesons is
studied in the context of nonet symmetry.  We have found that it is badly
broken
in the PP channels and in the P sector of the PV channels as expected from the
non-ideal mixing of the $\eta$ and the $\eta'$.  In the VV channels, it is also
found that nonet symmetry does not describe the data well. We have found that
this discrepancy cannot be attributed entirely to SU(3) breaking at the usual
level of 20--30\%. At least one, or both, of nonet and SU(3) symmetry must be
very badly broken. The possibility of resolving the problem in the future is
also discussed.
\end{quote}

\newpage 

\pagestyle{plain}

\sect{1.~~Introduction}
The original motivation for nonet symmetry \Ref{1} is the near `ideal' mixing
of
the $\omega$ and $\phi$ mesons and the extension of nonet symmetry to the
pseudoscalar sector goes back to the original study of charmed meson decays of
ref.~2. In this paper, we investigate the validity of and nonet symmetry in
Cabibbo-allowed two-body decays of charmed mesons (D) into pseudoscalar (P) and
vector (V) mesons.

In an earlier paper \Ref{3}, one of us has already shown that nonet symmetry
does not work in the decays of charmed mesons to two pseudoscalars.  We repeat
the analysis here in light of more recent data and we also show that it does
not
work for D $\to$ PV, possibly through a breakdown in the P sector. As a further
test, we will look at D $\to$ VV where it is expected to work. However, we find
that it does not! An alternative explanation using the breakdown of SU(3)
symmetry is also explored. The present status of experimental data does not yet
allow us to draw any definite conclusions.

The discussion of charm decay in the context of flavor SU(3) \Ref{4} and nonet
symmetry \Ref{5} has been treated in great detail in the literature.
Here we adopt the notations of ref.~[\,3\,]. The tensor structure of the
Hamiltonian governing Cabibbo-allowed decays gives rise to two representations,
a \R{15} and a \R{6*}. As before, we construct these from the two final state
nonets and the charm meson triplet. The incorporation of the singlet into the
octet to form the nonet does not alter the Clebsch-Gordon series of $\R{8}
\otimes \R{8} = \R{27} + \R{10} + \R{10*} + \R{8} + \R{8} + \R{1}$.  We will
label the reduced matrix elements obtained from the \R{27} as $T$, those from
the \R{10} and \R{10*} as $D$ and those from the symmetric and antisymmetric
combinations of the two \R{8}'s as $S$ and $A$ respectively.  Finally, the
overall representation will be denoted by a subscript so that $T_{15}$ would be
the reduced matrix element of the \R{15} obtained from $\R{27} \otimes \R{3}$.

The $D$'s and the $A$'s are antisymmetric under the exchange of the final state
mesons so that they do not contribute to PP and VV channels. It is convenient
to
introduce the combinations $S_{\pm} = S_{15} \pm S_6$ and $A_{\pm} = A_{15} \pm
A_6$ because each of these occurs only in the decays of either the $D^0$ meson
or the $D_s$ meson but not both.  The decay amplitudes are summarized in
tables~1, 2, and 3 along with the relevant phase space factors and
experimentally measured branching ratios.

When nonet symmetry is broken, amplitudes involving the singlets are no longer
related to those involving only members of the octets. To parametrize the
extent
of the breaking, we  keep the original amplitudes under nonet symmetry and
introduce a new amplitude $B$ for each channel involving a singlet.  Two such
amplitudes are used in each of tables~1 and 3 for the discussion of the PP and
VV channels.

For computational purposes, we will define all amplitudes in terms of branching
ratios expressed in percent by the following relation
$$
  B(D_i \to XY) = \frac{|A(D_i \to XY)|^2 \times \hbox{phase space factor}}
                       {\Gamma(D_i)/\Gamma(D^0)}~.                   \eqno{(1)}
$$
To allow for final state interactions, we have taken all amplitudes to be
complex. To solve for both the real and imaginary parts of one, we will need
two
branching ratios. Also, because of the quadratic nature of eq.~(1), there will
in general be a two-fold ambiguity in each solution.

Error analysis in this kind of calculation is rather complicated, but we try to
take into account correlations as much as possible.  Since we are always
comparing amplitudes or branching ratios, the total widths that enter eq.~(1)
will appear in all quantities of interest. The uncertainties in the total
widths
have been set to zero in order to avoid artificially inflating the errors
involved in such comparisons.  We treat uncertainties in all measured branching
ratios as independent, but will try to keep track of their propagation into
various derived quantities correctly.  In cases where a common normalization is
used, such as in $D_s$ decays, the ratios instead of the absolute branching
ratios will be treated as independent.  For comparisons involving only $D_s$
modes, the uncertainty in the common normalization, $B_{\phi\pi^+}$, will again
be dropped.

We will discuss the decay of charmed mesons into PP, PV and VV channels
separately in sections 2, 3 and 4. Section 5 will be devoted to the study of
possible SU(3) breaking in VV and PP channels. We conclude with a brief summary
in section~6.

\sect{2.~~D $\to$ PP}
Table 1 summarizes all relevent information about the PP channels.  Almost all
Cabibbo-allowed modes are very well measured, especially in the $D^0$ sector.
Even in the presence of the nonet breaking amplitude $B_+$, we have enough data
to solve for all other amplitudes. The $D^+$ branching ratio gives us directly
the magnitude of $T_{15}$. Taking $T_{15}$ to be real, we can then solve for
$S_+$ from mode 2 and 3. To solve for $B_+$, we need to express the
$\eta_8,~\eta_1$ amplitudes in terms of those of $\eta,~\eta'$.  We do this by
using a pseudoscalar mixing angle \Ref{11} of $-20\degree$ and the solutions
are    $$
\begin{array}{l}
  5\,T_{15} = 1.10 \pm 0.08~, \\
  S_+ = (0.10 \pm 0.23) + i (1.93 \pm 0.07)~, \\
  B_+ = \left\{ \begin{array}{l}
                      +(1.86 \pm 0.23) - i(1.61 \pm 0.49) \\
                      -(1.94 \pm 0.36) - i(2.70 \pm 0.41)~.
                    \end{array} \right.
\end{array}
$$
Another set of equally good solutions can be obtained by a reflection about the
real axis. The large size of $B_+$ clearly indicates that nonet symmetry in
this
sector of charm decay is badly broken.

In the $D_s$ sector there are two new amplitudes, $S_-$ and $B_-$, but only
three extra branching ratios.  We do not have enough information to solve for
everything, but there exist triangular sum rules which impose constraints on
the
amplitudes.

To simplify the discussion, we set $B_- = 0$ and see if this would leads to any
contradictions. Table~1 gives us the following relations
$$
  \sqrt{3} A(\pi\eta_1) - \sqrt{6} A(\pi\eta_8) = 6 T_{15}~,         \eqno{(2)}
$$ $$
  \frac{5}{2} \sqrt{3} A(\pi\eta_1) - \sqrt{6} A(\pi\eta_8) = 3A(K^+\Kbar^0)~.
                                                                     \eqno{(3)}
$$
Putting in the pseudoscalar mixing angle $\theta$ to convert $\eta_{1,8}$ into
$\eta,~\eta'$, we obtain from eq.~(2):
$${\renewcommand{\arraycolsep}{0.1em}
\begin{array}{l c c l}
  & (\cos\theta-\sqrt{2}\sin\theta) A(\pi\eta')
  & -~(\sqrt{2}\cos\theta+\sin\theta) A(\pi\eta) = & 2\sqrt{3} T_{15} \\
  \theta = -20\degree: & 2.87 \pm 0.44 & 1.21 \pm 0.20 & 0.76 \pm 0.05 \\
  \theta = -10\degree: & 2.48 \pm 0.38 & 1.50 \pm 0.25 \\
\end{array}
}$$
The numbers under each term are the values of that term evaluated with the
mixing angles indicated.  For $\theta=-20\degree$, we have $2.87 - 1.21 = 1.66
>
0.76$.  Even if we take into account the errors in each term, there is no
choice of phase in which the three amplitudes can form a closed triangle. This
is strong evidence that $B_-$ has to be nonzero. For $\theta=-10\degree$, the
three amplitudes are barely consistent with $B_-=0$, if the errors are
stretched
to their limits.

The corresponding results for eq.~(3) are
$${\renewcommand{\arraycolsep}{0.1em}
\begin{array}{lc c l}
  & (\frac{5}{2}\cos\theta - \sqrt{2}\sin\theta) A(\pi\eta')
  & -~(\sqrt{2}\cos\theta + \frac{5}{2}\sin\theta) A(\pi\eta) =
  & \sqrt{3}A(K^+\Kbar^0) \\
  \theta = -20\degree: & 3.57 \pm 0.28 & 0.36 \pm 0.04 & 1.84 \pm 0.16 \\
  \theta = -10\degree: & 3.41 \pm 0.27 & 0.74 \pm 0.07 \\
\end{array}}                                                         \eqno{(4)}
$$
Here, the common factor $B_{\phi\pi^+} \Gamma(D_s)/\Gamma(D^0)$ in the three
amplitudes has been taken out so that the listed errors can be treated as more
or less independent.  If we take the difference between the two terms on the
left, we obtain $3.21 \pm 0.28$ for $\theta = -20\degree$ and $2.67 \pm 0.28$
for $\theta = -10\degree$. To form a triangle, these have to be less than $1.84
\pm 0.16$, which is certainty not true.  Thus we have shown once again that
prediction from nonet symmetry does not agree with data.

\sect{3.~~D $\to$ PV}
The sheer number of independent amplitudes in the PV channels make the analysis
much more complicated. Because of the abundance of data, all quantities in the
$D^0$ modes can be solved in terms of the relative phase between the two $D^+$
amplitudes.  However, as mentioned earlier, there  are discrete ambiguities in
the solutions, and this makes precise predictions very difficult.  So far, we
have found no inconsistency between the data and amplitudes shown in table~2.
We will see below that this is not the case with $D_s$ decays.

The experimental situation in the $D_s$ modes is somewhat less developed, but
there are three sum rules that we can use.  We will try to determine the effect
they have on the amplitude $S_-$.  First of all, from the $K K^*$ channels we
have
$$
  A(K^+\Kbar^{*0}) + A(\Kbar^0K^{*+}) =
  2S_- + \frac{2}{5}A(\Kbar^0\rho^+) + \frac{2}{5}A(\pi^+\Kbar^{*0})~.
$$
Putting in numerical values for the amplitudes gives
$$
  |S_-| \le 3.64 \pm 0.33~.                                          \eqno{(5)}
$$
{}From the $\pi^+$-$\omega,\phi$ sector, with only an upper limit for the
$\omega$
mode, we have
$$ \begin{array}{l l}
         & 2\,S_- = \sqrt{2}A(\pi^+ \omega) + A(\pi^+ \phi)~, \\
  \imply & 0.27 \pm 0.17 \le |S_-| \le 2.33 \pm 0.17~.
\end{array}                                                          \eqno{(6)}
$$
Finally, expressing $\eta_1$ in terms of $\eta,~\eta'$ gives
$$
\begin{array}{l c}
 &\frac{2}{\sqrt{3}}S_- = A(\eta'\rho^+)\cos\theta - A(\eta\rho^+)\sin\theta\\
  \theta=-20\degree: & 6.02 \pm 1.13 \le |S_-| \le 8.54 \pm 1.38 \\
  \theta=-10\degree: & 6.99 \pm 1.25 \le |S_-| \le 8.27 \pm 1.37
\end{array}                                                          \eqno{(7)}
$$
The relations (5) and (6) are compatible with each other but not with (7).
The natural conclusion, of course, would be that there is large nonet symmetry
breaking in the $\eta$-$\eta'$ sector but none in the $\phi$-$\omega$ sector.

To test the idea of nonet symmetry in the $\phi$-$\omega$ sector we look at
pure VV decays in the next section.

\sect{4.~~D $\to$ VV}
In VV channels partial waves with $L=0,~1,~2$ can all contribute.
Since the available phase space is generally small and the phase space factor
depends on the center of mass momentum as $p^{2L+1}$, $s$-waves tend to
dominate.  This is well supported by data of the $\rho K^*$ modes \Ref{9}.
Keeping only $s$-waves makes the SU(3) amplitudes of the VV modes look very
similar to those of the PP modes with the exception that $D^0 \to \phi
\Kbar^{*0}$ is kinematically forbidden. This means that we will not have enough
information to determine both the phase and modulus of the $B_+$ amplitude.

Table 3 summarizes the situations. It is obvious that we can still determine
$T_{15}$ and $S_+$ from the three $\rho K^*$ modes. By assuming $B_+$ to be
zero, we can make a prediction for the $\omega K^*$ mode:
$$
  B(D^0 \to \omega \Kbar^{*0}) = 2.6 \pm 2.0~.                       \eqno{(8)}
$$
This prediction will, of course, change as the quality of the data improves.
If, in the future, this branching ratio turns out to be significantly different
from the predicted value, we will have to conclude that $B_+$ has to be big.
We now turn to the $D_s$ modes.

In the absence of nonet symmetry breaking, table~3 tells us that $A_1$, the
$D^+$ amplitude, would be 2.5 times as big as $A_8$, the $\phi\rho^+$
amplitude. Phase space for the two cases are comparable; with the lifetime of
the $D^+$ being 2.5 times as big as that of the $D_s$, this translates directly
into a factor of more than fifteen for the branching ratios. This is in serious
contradiction with data and we are left with the disturbing fact that $B_-$ is
indeed big---something that has not been born out by the PV analysis.  Our
reluctance to give up nonet symmetry for the vector mesons drives us to look
for
other explanations. In the next section we will investigate the possibility
that
a small SU(3) breaking may be responsible for the discrepancy.

\sect{5.~~SU(3) breaking}
If we assume that the comparatively large $s$ quark mass to be solely
responsible for the breaking of flavor SU(3) in strong interactions, the
effective Hamiltonian will transform as an octet \Ref{12}. Coupled to the
original SU(3) conserving piece, it gives us in the PP and VV cases two extra
representations \Ref{13}: \R{42*} and \R{24}, both coming from the \R{27} of
$\R{8} \otimes \R{8}$. They appear in table 4 as $T_{42}$ and $T_{24}$.

In the wake of SU(3) breaking, the situation has become much more complicated,
but two things remain unchanged. The first is the isospin relation of $A_1$,
$A_2$, and $A_3$; isospin is obviously still a good symmetry by design. The
second is that the two amplitudes involving the singlet $\omega_1$ are
unaffected. Unfortunately, one of these involves the kinematically forbidden
$\phi K^*$ mode. The other one also gives us the following relation
$$
  5B_- = \sqrt{2}A_7 + 3A_8 - 2A_6~.
$$
This relation holds regardless of whether there is SU(3) breaking or not. If we
can show that the three amplitudes on the right hand side do not form a closed
triangle, then we definitely have nonet symmetry breaking; the contrary,
unfortunately, is not true.  Presently, there is no measurement of the
$\rho^+\omega$ branching ratio.  We can always let $B_- = 0$ and look forward
to
its value allowable under nonet symmetry:
$$
  \Bigl| 3|A_8|-2|A_6| \Bigr| \le \sqrt{2}|A_7| \le 3|A_8|+2|A_6|   \eqno{(9)}
$$
gives
 $$
  1.7\pm3.9 \le \frac{B(D_s\to\rho^+\omega)}{B(D_s\to\pi^+\phi)} \le 86\pm27~.
$$
The large upper limit is mainly due to the much larger phase space of the $\rho
\omega$ mode compared to those of the $K^*K^*$ and the $\rho\phi$ modes. With
the improvement of data, the lower limit may prove to be useful if the central
values of the present branching ratios remain unchanged. Returning to $A_1$ and
$A_8$ and judging from the expressions in table~4, it is inevitable that at
least one of $B_-$, $T_{24}$, and $T_{42}$ must be big.

Relation (9) can also be used for the PP channels.  The corresponding relation
for the pseudoscalar case is exactly relation (4) we obtained earlier.
Therefore
in the case of D $\to$ PP we have effectively shown that nonet symmetry is not
obeyed regardless of whether there is SU(3) breaking or not.

\sect{6.~~Summary}
We have shown in section 2 that nonet symmetry is badly broken in the PP
channels. Even with the help of SU(3) breaking, one cannot evade this
inevitable
consequence. The situation with the PV channels is less clear. It seems that
nonet symmetry may still be good in the vector sector but not in the
pseudoscalar sector. Also, we have not found any evidence for SU(3) breaking
there. Compared with these two cases, the situation in the VV channels is much
more confusing. First of all, we have shown by comparing $D^+ \to
\Kbar^{*0}\rho^+$ and $D_s \to \phi\rho^+$ that at least one of the three
symmetry breaking amplitudes must be large. Secondly, we have not been able to
rule out the possibility of having large SU(3) breaking effects. In fact, it
will not be possible for us to completely rule out the breakdown of either
SU(3)
or nonet symmetry by studying Cabibbo-allowed decays alone because there are
simply too many free parameters. However, if nature cooperates, we may be able
to confirm explicitly the breakdown of nonet symmetry in the vector sector. The
prospect of seeing this in the VV channels relies on the overall improvement of
data and the observation of two channels [\,relations (8) and (9)\,] both
having
the $\omega$ in their final states.

Finally, there is also the possibility of contributions from higher partial
waves.  In particular, $p$-wave contributions will lead to new SU(3) amplitudes
so that there will be as many independent amplitudes in the VV channels as
there are in the PV channels.  Though unlikely to be the case, this is an issue
that can be resolved experimentally.  So far, it has not been supported by
data \Ref{9}.

This work was supported in part by the U. S. Department of Energy grant
DE-FG05-92ER40691.

\newpage 

\def \jour#1,#2,#3,#4'{#1  #2 (19#4) #3}

\def \etal{et al.}

\def \PRD{Phys.\ Rev.\ D}
\def \PRL{Phys.\ Rev.\ Lett.}
\def \NP {Nucl.\ Phys.}
\def \PL {Phys.\ Lett.}
\def \ZP {Z.\ Phys.}

\setlength{\parindent}{0em}
\sect{Reference}

\renewcommand{\tabcolsep}{0.1em}
\begin{tabular}{r p{5.7in}}
\Ref{1} & 
S. Okubo, \jour\PL,5,165,63'. \\
\Ref{2} & 
M. B. Einhorn and C. Quigg, \jour\PRD,12,2015,75'. \\
\Ref{3} & 
S. P. Rosen, \jour\PRD,41,303,90'. \\
\Ref{4} & 
R. L. Kingsley, S. B. Treiman, F. Wilczek, and A. Zee, \jour\PRD,11,1919,75';
L.-L. Chau and F. Wilczek, \jour\PRL,43,816,79';
G. Altareli, N. Cabibbo, and L. Maiani, \jour\NP,B88,285,75';
M. B. Voloshin, \etal, \jour Pis'ma Zh.\ Eksp.\ Teor.\ Fiz.,21,403,75'
   [\jour JETP Lett.,21,183,75'.] \\
\Ref{5} & 
C. Quigg, \jour\ZP,C4,55,80';
S. P. Rosen, \jour\PL,B218,353,89';\jour\PL,B228,525,89';
   \jour\PRD,39,1349,89';
Y. Kohara, \jour\PL,B228,523,89';
A. N. Kamal and R. C. Verma, \jour\PRD,35,3515,87';\jour\PRD,43,829,91'. \\
\Ref{6} & 
J. J. Hernandez, \etal, \jour\PL,B239,1,90'. \\
\Ref{7} & 
obtained by combining results of the following three experiments:
(1) ARGUS, P.~Karchin, in: Proc.\ 1989 Int.\ Symp.\ on Lepton and Photon
   Interaction at High Energies, Stanford, ed.\ M. Riordan, World
   Scientific, Singapore (1990) p.~105;
(2) MARK III, J. Labs, \etal, \jour\NP,A527,753,91';
(3) CLEO, W.-M. Yao, in: The Vancouver Meeting, Particles and Fields '91,
   ed.\ D. Axen, D. Bryman, and M. Comyn, World Scientific, Singapore
   (1992) p.~416. \\
\Ref{8} & 
M. Daoudi, \etal, CLEO Collaboration report CLEO-91-9 (same as CLNS-91-1108),
   Sept.~1991. \\
\Ref{9} & 
D. Coffman, \etal\ (MARK III Collab.,) \jour\PRD,45,2196,92'. \\
\Ref{10} & 
S. Barlag, \etal\ (ACCMOR Collab.,) \jour\ZP,C48,29,90'. \\
\Ref{11} & 
J. F. Donoghue, B. R. Holstein, and Y. C. R. Lin, \jour\PRL,55,2766,85';
F. J. Gilman and R. Kauffman, \jour\PRD,36,2761,87'.
See also Ref.~6, p.~III63. \\
\Ref{12} & 
M. Gell-mann, Caltech Synch.\ Lab.\ Report No.\ CTSL-20, reprinted in M.
Gell-mann and Y. Ne'eman, The Eight fold Way, Benjamin, New York (1964);
S. Okubo, \jour Prog. Theor. Phys.,27,949,62'. \\
\Ref{13} & 
M. Savage, \jour\PL,B257,414,91'; \jour\PL,B259,135,91'.
\end{tabular}

\end{document}